\documentstyle[prd,aps,epsfig]{revtex}
 \tighten
\begin{document}
\draft
\twocolumn[\hsize\textwidth\columnwidth\hsize\csname 
@twocolumnfalse\endcsname
\title{CAN HIGH ENERGY COSMIC RAYS BE VORTONS~?}
\author{Silvano BONAZZOLA and Patrick PETER}
\address{D\'epartement d'Astrophysique Relativiste et de Cosmologie,\\
Observatoire de Paris-Meudon, UPR 176, CNRS, 92195 Meudon (France).}
\date{\today} 
\maketitle
\begin{abstract}
A simple model is exhibited in which the remnant density of charged
vortons is used to provide candidates for explaining the observed
ultra high energy cosmic rays (above $10^{20}$~eV). These vortons
would be accelerated in active galaxies and propagated through
intergalactic medium with negligible losses of energy. The expected
number density of observable events is shown to be consistent with
extrapolation of the observations. The spectrum is predicted to be
spatially isotropic while its shape is that of an atomic
excitation-ionisation, i.e. with a few peaks followed by a continuum;
there is also an energy threshold below which no vorton is visible.
\end{abstract}
\vskip2pc]

\section*{Introduction}

The problem of very high energy cosmic rays~\cite{bird} is still an
open one~: most models based on reasonable astrophysical assumptions
seem to indicate a likely maximum value for the energy of any kind of
emitted particle at most of $\sim 10^{19}$~eV~\cite{sigl} and the
existence of the microwave background makes it impossible for a proton
say to propagate with such energy on scales much larger than a few
tens of Mpc~; this is the celebrated Greisen-Zatsepin-Kuz'min (GZK)
cutoff~\cite{GRZ}. An observation dated October 1991 by the Fly's Eye
detector showed evidence for a cosmic ray at an energy of $(3.2\pm
0.6) 10^{20}$~eV~\cite{bird} and a few others reported showers above
$10^{20}$~eV~\cite{others} (and well above the GZK cutoff). There does
not seem to date to be any completely satisfactory suggestion for
explaining these observations (see however Ref.~\cite{CenA}), and it is
the purpose of this paper to propose such a possible explanation that
has been overlooked, namely that cosmic rays may consist of (as
opposed to originate in) topological defects (as was first proposed in
Ref.~\cite{magmon}).

A fairly common (neither confirmed nor excluded) prediction of
particle physics models at energies beyond that of the electroweak
scale is the existence of topological
defects~\cite{kibble,book}. Among these, only cosmic strings seem to
be consistent with cosmological constraints (with the possible
exception of textures which, as non-localized objects, are not
considered here). In particular, strings have been proposed as seeds
for galaxy formation as well as sources for the microwave background
fluctuations observed by COBE. This requires that the strings appear
at the Grand Unified (GUT) scale, a scenario that may breakdown in the
case suggested by Witten~\cite{witten} in which they would be endowed
with superconducting properties.  In the latter case, string loops can
form equilibrium configurations called vortons~\cite{vorton,ring},
whose stability is currently under
investigation~\cite{stabgen,stabwit}, that would easily overfill the
Universe~\cite{vorton,ring}, thus provoking an unobserved cosmological
catastrophe. It has been suggested, as a possible way out of this
problem, that cosmic strings might have appeared at a much lower
symmetry breaking scale, estimated in the range $\sim
10^4-10^9$~GeV~\cite{ring,ring2} (or at least that the formation of
superconducting currents may have been postponed until this stage).

A second reason to consider such comparatively low energy cosmic
strings instead of the more popular GUT strings is that the latter
have been shown~\cite{gill} to yield a negligible flux of cosmic ray
types events as compared to what is currently observed, whatever the
mechanism involved. This is mostly based on the idea that, a string
being a topologically stable object, emission of very high energy
particles must proceed via some removal of the topological stability,
which for a GUT string implies either extremely highly energetic
phenomena (and we're back to the previous problem of reaching the
required high energies) or relatively rare situations (cusp
evaporation, intercommutation and loop collapse). Adding up both
cases, the expected flux turns out to be very tiny (roughly $10^{-10}$
times that observed). Our aim here is to show that it is possible in
principle to consider high energy rays as cosmic string loops
themselves if the scale is not that of GUT but rather much below that
scale. This, as we shall see, gives a flux that is comparable with
observations.

The paper is organized as follows: in Section~\ref{phys}, we recall
the basic facts about vortons and how one may expect them to interact
with ordinary particles. Then, Section~\ref{eff} is devoted to the
interaction of a vorton with a proton at rest in order to estimate the
order of magnitude of the various phenomena involved in the detection
of a $10^{20}$~eV event, including the probability that a vorton
interacts with the atmosphere. It is predicted that the typical
expected spectrum form should resemble that of an
excitation-ionisation spectrum, namely it should consist of lines
followed by a continuum. Besides, since the interaction probability is
found to be quite low (the cross-section is typical of neutrino-hadron
interactions at the same energies~\cite{nus}), we predict important
horizontal showers.  In Section~\ref{sec:II}, we present a plausible
acceleration mechanism for vortons, which turns out to be much simpler
than most of the standard acceleration mechanisms for protons
(Fermi-like mechanisms), without sharing their drawbacks.  We then
discuss propagation in Section~\ref{sec:III} where it is shown that no
GZK cutoff is to be expected for vortons.  It is then argued that this
model implies that the most energetic events should be distributed
isotropically. Finally we summarize our findings in the conclusion
where we also compare the expected properties of these rays with the
anticipated detection capabilities of the future Auger
observatory~\cite{auger}.

\section{The physics of vortons}\label{phys}

The objects we shall now consider are
vortons~\cite{vorton,ring,ring2}, namely loops of cosmic strings
endowed with superconducting properties and stabilized by a
current. They arise in symmetry breaking theories above the
electroweak scale for which the vacuum manifold is not simply
connected. In other words, if the high energy vacuum is invariant under
the transformation of a symmetry group $G$ and the low energy vacuum
under that of the group $H$, the necessary and sufficient condition
for the existence of cosmic strings is that the first homotopy group
$\pi_1$ of the quotient $G/H$ be nontrivial, $\pi_1 (G/H)\not\sim \{ 0
\}$. A typical example is the scheme $U(1)\to \{ 0\}$, which is also
the scheme at work in the Landau-Ginzburg model for
superconductivity~\cite{GL}, and the strings are the corresponding
vortices, much studied experimentally~\cite{exp}. The vortices appear
as infinite or in the form of closed loops which decay via emission of
gravitational radiation. The symmetry breaking is achieved thanks to a
Higgs field $\Phi$ which in turn can be coupled to other fields,
particularly charged ones, which we shall generically write as
$\Sigma$. Here we consider only the case where this charge carrier is
a scalar particle, noting anyway that because of the formal
equivalence between bosons and fermions in the two-dimensional
worldsheet generated by the string, one expects similar behaviors for
any kind of particle. Besides, as was shown to be sufficient in
previous work\cite{neutral}, we shall only consider the coupling of
$\Sigma$ with the electromagnetic field as a perturbative effect not
modifying much the overall dynamics of the vorton.

Let us here summarize the basic properties of vortons~\cite{ring2} as
we expect to observe them. They can be characterized by essentially
two integer numbers, namely a topological one, $N$ say, which
specifies the winding of the current carrier phase around the loop,
and a dynamically conserved number, $Z$, related in the
charge-coupled case (to which the present analysis is restricted) with
the total amount of electric charge $Q$ that it holds through $Q=Ze$,
with $e$ the electron electric charge. Note that both $N$ and $Z$ can
be positive or negative, although for $N$ it is just a matter of
convention, while in the case of $Z$, the sign is important when there
is an external interaction (as in the case of acceleration for
instance). Both these numbers are conserved at a classical level and
one expects them to be of the same order of magnitude $Z\simeq N \sim
100$~\cite{ring2}. From these, on constructs the loop angular momentum
as $J=ZN\sim 10^4$: although a bit hypothetical and thus often seen as
``exotic'', vortons can actually be considered as classical objects.

Since it can be shown that electromagnetic self-coupling is mostly
negligible~\cite{enon0}, one can evaluate the total mass-energy of the
vorton simply in terms of its characteristic radius $r_{_V}$, its energy
per unit length $U\sim m^2$ and tension $T\sim m^2$ as~\cite{14}
\begin{equation} M_{_V} = 2\pi r_{_V} (U+T) \sim m^2 r_{_V},
\end{equation}
with $m$ the energy scale at which the strings are formed (essentially
the string forming Higgs mass). Moreover, we shall assume in what
follows that the current carrier mass scale $m_\sigma$ is also of
order $m$ so that we can keep only one energy scale. Note however that
the following analysis can be easily generalized for two different
mass scales since in practice it is the current scale that is relevant
in most calculations. Then, knowing the angular momentum to be given
by~\cite{14} $J^2 = UT r_{_V} /2\pi$ permits to calculate the
characteristic vorton circumference as
\begin{equation} 2\pi r_{_V} = (2\pi)^{1/2} |NZ|^{1/2} (UT)^{-1/4}
\sim Z /m \label{length}\end{equation}
with a corresponding mass
\begin{equation} M_{_V} \sim Z m. \end{equation}
Moreover, the mass scale is constrained: depending on various
assumptions about the string network evolution and the rate of loop
formation as well as the probability that arbitrarily shaped loops end
up in vorton states, it can be shown that, if the vortons are
stable~\cite{stabgen,stabwit}, then in order to avoid a cosmological
mass excess ($\Omega_{_V}<1$), the condition
\begin{equation} m < 10^9 \ \hbox{GeV},\label{constraint}
\end{equation}
must be satisfied,

Having discussed the basic properties of vortons, let us now turn to a
rough evaluation of the typical mass scale expected for $m$ if those
vortons were to be seen as cosmic rays, developing air showers. As we
shall see in the following section, interaction between a vorton and
whatever other particle occurs mainly via inelastic scattering
resulting in the extraction of a trapped $\Sigma$ particle. In other
words, the current flowing along the string loop can be seen as a
bunch of bound states which can be excited provided the interaction
energy is large enough. We therefore conclude at the existence of an
energy threshold above which the typical expected spectrum should
change qualitatively. Besides, and we now come to the second firm
prediction of the model, since those particles form bound states, we
expect them to show up in the form of a line spectrum, bound states
being always quantized; this will be shown on a specific vorton model
in section~\ref{eff}. For the time being, what really matters is the
existence of bound state energies the order of magnitude of which we
shall now attempt to evaluate.

Let $\Delta E$ be the variation of energy between two energy levels in
the vorton, calculated in its rest frame, and $\gamma$ its Lorentz
factor in the rest frame of the particle it interacts with (recall we
are at the end interested in air showers occurring in the atmosphere
where the particles interacting with the vorton, namely essentially
quarks composing protons and neutrons, are supposed to be at
rest). The energy $\epsilon$ at which the interaction is then seen is
obtained by transforming back to the particle's rest frame,
\begin{equation} \epsilon \sim \gamma \Delta E.\label{dE1}
\end{equation}
Denoting by $\widetilde m$ the particle's mass and requiring the
interaction to actually take place gives
\begin{equation} \gamma\widetilde m\sim \Delta E,\label{dE2}
\end{equation}
so that altogether, Eqs.~(\ref{dE1}) and (\ref{dE2}) imply
\begin{equation} \gamma^2 \sim {\epsilon \over \widetilde m},
\label{gamma2}\end{equation}
from which the characteristic cosmic string scale can be deduced in
the following way. From the fact that the angular momentum of the
vorton is $J\sim Z^2$, it is seen that the density of states scales
like $Z^{-2}$. Therefore, $\Delta E \sim m /Z^2$, so that using
(\ref{dE2}) and (\ref{gamma2}) yields
\begin{equation} m\sim Z^2\sqrt{\epsilon \widetilde m}.
\label{m}\end{equation}
We shall now use these relations together with the observations that
have been realized on cosmic rays to normalize the energy levels..

Let us apply this evaluation to ultra high energy cosmic
rays~\cite{bird} so that the characteristic observed energy is
normalized to $\epsilon \sim 10^{20}$eV. The particles vortons
interact with are essentially quarks composing hadrons, so the mass
$\widetilde m$ can be taken to be that of the proton (at this level of
approximation, the mass difference between a quark and a proton is
negligible). With these numbers in mind, Eq.~(\ref{gamma2}) transforms
to $\gamma^2\sim 10^{11}$ and the cosmic string scale (\ref{m}) $m\sim
10^9$~GeV which, surprisingly enough, turns out to be right at the
closure limit given by Eq.~(\ref{constraint}). It now remains to
calculate the effective cross-section between a vorton and a quark in
order to know the expected flux and check that it is compatible with
the current observational limits.

\section{Cross-Section Vorton-hadron}\label{eff}

There is no privileged model for cosmic strings, and therefore neither
is there for describing vortons. Thus, we can at best evaluate rough
orders of magnitude for the cross section we are looking for. There
are however various levels of approximation at which a vorton can be
described. As mentioned earlier, the characteristic length scale
associated with a vorton configuration is expected to be a hundred
times larger than its thickness, so the most obvious description of a
vorton is that of a classical circle. This is unfortunately of
absolutely no use when one is looking for the trapped $\Sigma$
particles along the loop. Hence, as a second level of approximation,
we shall consider a vorton to be a torus like configuration in which
the field $\Sigma$ feels a confining potential. This is beyond the
scope of the present calculation and is left for further
investigation~\cite{preview}. So let us describe a vorton as a sphere
of radius $R$ where in fact the mass of $\Sigma$ is a radially
dependent function. In order to try and recover the circular geometry,
the confined field, expanded as it should in the eigenvectors of
angular momentum (spherical harmonics), will be described only by
those high values of the total angular momentum (to take into account
the fact that $J\gg 1$) as well as its projection along a fixed axis.

The basis for the possible bound states of $\Sigma$ will therefore be
given as the set of quantized solutions of the Klein-Gordon equation
\begin{equation} (\Box + M^2 ) \Sigma =0,\label{KG}\end{equation}
where the mass $M=M(r)$.  The effective radius $R$ is adjusted in such
a way that the geometrical section of the corresponding sphere $\pi
R^2$ is that of the equivalent ring if it hits a particle face-on,
i.e. $\pi R^2=\pi [(r_{_V}+r_{_X})^2 - (r_{_V} - r_{_X})^2]$, or $R^2
= 4 r_{_V} r_{_X}$. In these formula, $r_{_V}$ stands as before for
the vorton radius, while $r_{_X}\sim m^{-1}$ is its thickness,
i.e. $R\sim Z^{1/2} m^{-1}$.

In the particular example which was used to calculate numerically the
curve on Fig.~1, the mass function in Eq.~(\ref{KG}) is taken as $M(r)
= m \Theta (R-r)$, and the trapped field is assumed the separated form
$\Sigma (x^\alpha )= u_{n\ell} (r) Y_{\ell {\cal M}}(\theta ,\phi)
\hbox{e}^{i\omega_{n\ell {\cal M}} t}$, with $Y_{\ell {\cal M}}(\theta
, \phi)$ a spherical harmonic and the radial function provides the
quantized energy levels for each value of the angular momentum by
imposing equality of the logarithmic derivative of the normalizable
solutions at $r=R$ (in the simplified model we investigate, those are
spherical Bessel and Hankel functions of the first kind). As discussed
above, we shall restrict our attention to ${\cal M}\sim \ell\sim
Z^2\sim 10^4$ in order to take into account, at least partially, the
effectively circular (as opposed to spherical) symmetry.  For the
actual calculations of effective cross-sections, we shall adopt the
same normalization convention for all the states, i.e. the bound and
free states (including hadron states) are covariantly normalized at
$2E$ particles in the volume $V$ for a state of energy $E$
($\omega_{n\ell {\cal M}}$ in the case of a bound state).

We now calculate the cross section $\sigma_{_V}$ for an incident
electromagnetically charged particle, a hadron say for definiteness,
in the frame of the loop, to interact inelastically with a vorton. We
consider the situation in the rest frame of the loop in which the
incident hadron with momentum $p_{in}=(E_1,{\bf p}_1)$ hits the
$\Sigma$ particle bound state characterized by its angular quantum
number $\ell_i$ and bound energy $\omega_1$ to yield an outgoing
hadron with momentum $p_{out} =(E_2,{\bf p}_2)$ and a final state for
the $\Sigma$ particle which could in principle be either another bound
state with quantum number $\ell_f$ or a free particle state with
momentum $k_f = (\omega_2, {\bf k})$; in the notation of the previous
section, one has $\Delta E=\omega_2-\omega_1$. In both cases, we
assume the decay rates $d\Gamma _{_I}$ of the resulting products,
namely either the excited state or the unstable particle itself, to be
much smaller than the energies of the particles so that the
cross-section for producing those unstable states effectively
factorizes as $d\sigma = d\sigma _{_I} d\Gamma _{_I}/\Gamma_{_I}$.
Integrating over the various decay possibilities ($\int
d\Gamma_{_I}=\Gamma_{_I}$) shows that it suffices to calculate only
the intermediate cross sections $d\sigma _{_I}$ and sum them up. Note
also that because our vorton model is very crude, we cannot expect
more than a rough order of magnitude for the cross section, a
detailed quantum examination of the same process being held for
another work~\cite{preview}. This is the reason why we now restrict
our attention to the ionisation process, including the possibility of
final bound state by considering various energy eigenvalues.

The interaction is assumed to be only electromagnetic. Hence, if
$\psi_{in}$ and $\psi_{out}$ are the wave functions of the hadron
under consideration before and after the collision, the 
hadronic current ${\cal J}^h$ we are interested in reads
\begin{equation} {\cal J}^h_\mu \equiv ie (\psi^*_{out} \partial _\mu 
\psi_{in}-(\partial _\mu \psi^*_{out}) \psi_{in} ),
\end{equation}
a similar definition holding for the $\Sigma$ current ${\cal
J}^\Sigma$ where the ``in'' state is a bound state, while the ``out''
state is a free $\Sigma$ particle. With the electromagnetic field
$A_\mu$, the total Hamiltonian for describing the collision is then
\begin{equation} H = {\cal J}^h_\alpha A^\alpha {\cal J}^\Sigma _\beta
A^\beta + H^\Sigma + H^h,\end{equation} where the self interaction
Hamiltonians $H^\Sigma$ and $H^h$ contain in principle all the
information about the internal structure on the interacting
particles. In practice, this means that we can work with eigenstates
of both these self-interacting Hamiltonians to take into account
exactly the vorton structure (this is supposedly done by calculating
the bound states) as well as the strong interaction effects binding
the quarks in the hadron together; thus, both protons and neutrons can
interact with a vorton, a point which is understandable by noting that
the vorton being much smaller than the hadron, it interacts
essentially with the quarks. Neglecting these corrections of order
unity, and working in the Lorentz gauge $\partial _\mu A^\mu =0$, the
amplitude for this process is easily calculated semi-classically as
(note we do not consider the fermionic degrees of freedom of the
hadrons)
\begin{equation} \langle p_{in},\ell_i |
p_{out}, k (\ell_f) \rangle = i \int d^4 x g^{\mu\nu} {\cal J}_\mu ^h
(x) q^{-2} {\cal J}_\nu ^\Sigma (x)\label{expand}\end{equation} where
$q=p_{out}-p_{in}$ is the exchanged momentum.  At this point, a
completely quantum field analysis could be made by expanding the
$\Sigma$ operator in bound and free particle creation and annihilation
operators and the hadron field in plane waves and imposing suitable
commutation relations (anticommutation to describe properly the hadron
states)~\cite{preview}. We recall that our purpose here is simply to
derive a rough order of magnitude estimate for the interaction cross
section, which explains why we perform only a semiclassical analysis
and neglect the fermionic behaviour of hadrons.  Use of
Eq.~(\ref{expand}) shall be made later for deriving the details of our
oversimplified spherical model, but for the time being, let us derive
the expected order of magnitude of the interaction cross-section.

\begin{figure}
\centering
\epsfig{figure=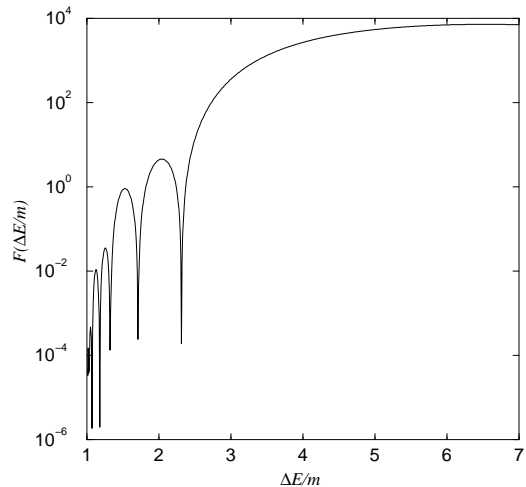, width=9cm}
\caption{Expected qualitative form of the cosmic ray spectrum for
vortons: it consists of a line spectrum followed by a continuum. This
function ${\cal F}(\Delta E/m)$ [see Eq.~(\ref{cs})] modulates the
cross-section and was calculated using 6 energy levels with $\ell
=10^4$, taking everything numerically into account.\label{Feps}}
\end{figure}

The total cross-section, being a sum over all the confined particles
of terms like Eq.~(\ref{expand}) squared is seen to have a factor $Z^2
e^4$, with $e$ the electromagnetic coupling constant. Then, on
dimensional grounds, it can be nothing but proportional to $(\Delta
E)^{-2}$. Thus, one ends up with
\begin{eqnarray} \sigma_{_V} &=& {Z^2 e^4\over (\Delta E)^2}{\cal F}
(\Delta E /m)\nonumber\\
&\simeq & 10^{-32} \left( {10^9 \ \hbox{GeV}\over m}\right)^2 Z_{100}^6
\ \hbox{cm}^2,\label{cs}\end{eqnarray}
where $Z_{100}\equiv Z/100$ and the function ${\cal F}$ is a
dimensionless quantity thats need a specific model to be evaluated.
In the spherical symmetry approximation and for a few energy levels,
it is exemplified on Fig.~\ref{Feps}.  The line and continuum spectrum
characteristic of bound state interactions is clearly shown on the
figure, as well as the existence of an energy threshold below which no
interaction takes place.

To conclude this section, we consider the probability that a vorton
interacts in the way discussed above in the atmosphere. Using the
values derived in the previous section for the mass and charge of the
vorton shows that the characteristic ionisation cross section is
typical of neutrino interaction at these energies~\cite{nus}. The
probability we are looking for is therefore, using $\rho_{Atm}\sim
10^3$~g$\cdot$cm$^{-2}$ for the mean atmospheric depth
\begin{equation} \alpha_{_V} = \sigma_{_V} \rho_{Atm} /m_{_P}\simeq 10^{-5}
-10^{-4} Z_{100}^6.\label{proba}\end{equation} This quantity we
shall use later to compute the expected flux of this type of events on
earth.

\section{Acceleration mechanism for vortons}\label{sec:II}

In this section and the following, we turn back to ordinary units in
which $c$ and $\hbar$ have their usual values.

Accelerating a particle of charge $Ze$ like a vorton to energies
larger than $10^{20}$~eV by means of an electric mechanism requires a
potential difference larger than $10^{20}/Z$~Volts. There are
basically two classes of astrophysical objects in which such potential
differences may be found, namely pulsars and accreting black holes
(BH). In fact, potential differences as high as $10^{18}$~Volts are
known to exist in the magnetosphere of young pulsars and in that of
the accretion disks around massive ($10^7~-~10^8 M_\odot$) Kerr BH.

Pulsars as sources of high energy vortons can be immediately ruled out
since vortons are not expected to be present at the surface of neutron
stars. Their mass is so high, as we have just seen, that they would
sink toward the center of the neutron star. We shall therefore for now
on consider the case of accreting BH, supposing those to be the power
engine for active galactic nuclei (AGN) radio galaxies and quasars
(QSO). The underlying idea is the following: radio jets, $X$ and
$\gamma$ rays emissions are due to the acceleration of electrons,
positrons and protons by electrostatic fields generated by a Blandford
type mechanism~\cite{blanford} near the horizon of the BH.  The model
is based on the assumption that a fraction $\alpha$ of the total
matter is made of vortons~\cite{ring2}, which behaves essentially as
cold dark matter. Therefore, one can expect their spatial distribution
to look like that of ordinary matter. The main idea is then that in an
accelerating object, whatever it is, the same fraction $\alpha$ of
vortons is accelerated by the same mechanism that works for the
protons.  However, in reasonable models, the total mass of the vorton
is $M_{_V}\sim Z m$ where $Z$ is the charge carried by the vorton and
$m$ the scale of symmetry breaking at which the strings were first
formed. As shown in Sec.~\ref{eff}, a Lorentz acceleration factor of
$10^5-10^6$ is actually sufficient to reproduce the data. This means
that a very basic acceleration by means of an electric field is enough
since at these energies, losses in radiation can be considered
negligible. In what follows, we shall justify the above model to show
that it encompasses no major difficulties.

It should first be clear, and we want to emphasize that point again,
that no exotic mechanism is required for the acceleration (even though
the accelerated particles themselves may be considered exotic).
Indeed, the acceleration mechanism that produces ultra-relativistic
jets in radio galaxies and $X$ and $\gamma$ rays in AGNs and QSOs is
enough to accelerate vortons to energies up to $10^{20}$~eV and
higher.  We want to point out that electrostatic fields are very
likely to be responsible for the acceleration of particles in jets of
radio galaxies and for the generation of high energy photons (with
energies exceeding 1~TeV). Although the origin of such electrostatic
fields is not completely understood yet, they may well originate
through the quite appealing Blandford--Znajecs
mechanism~\cite{blanford}.  Let us first describe shortly such a
mechanism.

\begin{figure}
\centering
\epsfig{figure=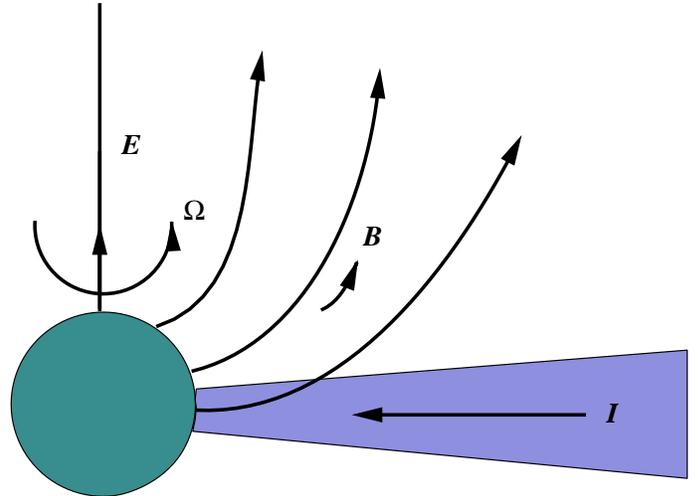, width=9cm}
\caption{The electromagnetic field configuration surrounding a Kerr BH
in the poloidal Blandford~Znajecs mechanism. The circle
represents the event horizon of the BH ($r=r_+$) and the shaded area
stands for the accretion disk.\label{F2sil}}
\end{figure}

Consider a magnetized rotating neutron star. Its surface is supposed
to be a perfect conductor, so that in the rest frame of the surface of
the neutron star, the tangential component $E_\theta$ of the
electrostatic field must vanish, a condition which, when written in
the inertial frame takes the form
\begin{equation} E_\theta - \big(
{{\bf \Omega} \times {\bf B}\over c} R_* \big)_\theta = 0,
\label{cond1}\end{equation}
where ${\bf \Omega}$, ${\bf B}$ and $R_*$ are respectively the angular
velocity, surface magnetic field and radius of the neutron star.  This
gives very energetic electric field lines along which charged
particles can be easily accelerated.

An analogous mechanism for generating electric field lines works for
Kerr BHs, with the unfortunate difference that a bare BH cannot have a
magnetic field, the latter being supplied by the accretion disk via a
dynamo mechanism. The surface of the BH behaves like a rotating
conductor with angular velocity $\Omega = \hat a c /M$, with $\hat a$
and $M$ the Kerr solution parameters somehow identifiable with the
total mass ($M$) and the angular momentum (for $0 < \hat a< 1$) of the
source. Now because of the boundary conditions on the surface of the
BH, an electrostatic field is created which yields a potential
difference given by~\cite{blanford1990}
\begin{equation} \Delta V = {\hat a B\over M} D^2, \label{DV}\end{equation}
where $B$ is the poloidal component of the magnetic field generated by
the accretion disk and $D$ the typical length scale of the
electrostatic field, $D\simeq M$, i.e., $\Delta V \simeq \hat aBM$ (see
Fig.~\ref{F2sil}).

The magnetic field is estimated to be $10^4$~gauss for the most active
galaxies. In fact, under the hypothesis that the magnetic energy
density is in equipartition equilibrium with the thermal energy of the
disk, one finds
\begin{equation} B=\left( {L_{_E}\over c}\right)^{1/2}
{c^2\over G M} \simeq 4\times 10^4
M_8^{-1/2}~\hbox{gauss},\label{B}\end{equation} 
where $L_{_E}$ is the Eddington luminosity
\begin{equation} L_{_E} = {4\pi G c M m_p\over \sigma_{_T}}
\simeq 10^{46} M_8~\hbox{erg}\cdot\hbox{s}^{-1},\end{equation} $m_p$
being the mass of the proton, $\sigma_{_T}$ the Thompson
cross-section~\cite{blanford} and $M_8=(M/10^8 M_\odot)$ being the BH
mass in units of $10^8$ solar masses. Taking into account
Eq.~(\ref{DV}), we then end up with a potential difference given by
\begin{equation} \Delta V \simeq 1.7\times 10^{20} \hat a
M_{8}^{1/2}~\hbox{Volts},\label{ii}\end{equation}
or, equivalently
\begin{equation} \Delta V =\sqrt{L\over c} \hat a \simeq 5\times 10^{19}
\hat a L_{45}~\hbox{Volts},\label{iii}\end{equation} $L_{45}$ being the
luminosity in units of $10^{45}$~erg/sec~\cite{blanford1990}.
Estimates of $B$ [Eq.~(\ref{B})] and $\Delta V$ [Eq.~(\ref{DV})] are
very rough and depend crucially on the conversion efficiency of
gravitational energy of the disk into electromagnetic energy. As
already stated, the parameter $\hat a$ is less than unity and is
related with the angular momentum $J$ of the BH through
\begin{equation} J = {\hat a\over c} G M^2,\end{equation}
where $M$ is now expressed in ordinary units. For old AGNs and QSOs,
$\hat a$ can be larger than $0.9$.

So high a potential difference is very efficient to accelerate
particles and can thus be the source for the electromagnetic energy of
AGNs, QSOs and radio galaxies. Similar powerful sources of
electromagnetic energy can be found in the magnetosphere of the
accretion disks and by radiation accelerated winds from the
disk. Interested readers will find details and useful references in
the review article~\cite{begel}

Hot spots in radio galaxy jets are also good candidates as
acceleration regions of very high energy particles. Shock waves in the
jets of radio galaxies are the main mechanism allowing acceleration of
protons at energies $\simeq 10^{20}$~eV, while protons in the near
zone of the BH cannot be accelerated at so high energies because of
the losses due to the unfavorable environment (synchrotron and Compton
losses)~\cite{nma}. Vortons on the other hand do not suffer from this
drawback since because of their high mass, they have a Lorentz factor
which is at most of the order of $10^6$ as we have seen above.

The synchrotron energy loss per unit length is
\begin{equation} {dE\over dx} = -{2\over 3} {(Ze)^2\over \rho^2}
\gamma^4,\end{equation}
with $\rho$ the curvature radius of the vorton trajectory along the
magnetic field lines. If we take $\rho$ equal to the Schwarzchild
radius $r_{_S}=2GM/c^2$ of the BH, we
obtain the total energy loss $\delta E$ as
\begin{eqnarray} \delta E &=& {(Ze)^2\over 3 G M} c^2 {\cal A} \gamma^4
\nonumber\\
&=& 5.2\times 10^{-33} Z^2  {\cal A} M^{-1}_8
\gamma^4 \ \hbox{erg}\end{eqnarray}
where ${\cal A}$ is the acceleration path expressed in Schwarzchild
units. For $Z=100$ and $\gamma =10^6$, the synchrotron losses are
$\sim 10^{-4}$~erg (with ${\cal A} =1$), a very small quantity indeed
with respect to the required final energy of $10^8$~erg.

A similar calculation can be performed for the Compton losses; let us
first compute the Thompson cross section $\sigma_{_V}^{_{Th}}$ of a
vorton:
\begin{eqnarray} \sigma_{_V}^{_{Th}} & = & 8\pi {(Ze)^4\over 3 M_{_V}^2 c^4}
\nonumber \\
& = & 1.6\times 10^{-45} Z_{100}^2 \left( {10^9 m_{_P} \over m}
\right) ^2 \ \hbox{cm}^2\end{eqnarray} 
(recall $M_{_V}=Zm$). It is now easy to compute the mean free path of
a vorton in the photon bath around a BH. For a given luminosity $L$,
the photon density $n_{ph}$ reads
\begin{equation} n_{ph}\simeq {L\over 4\pi R^2 c h\nu}\end{equation}
where $R$ is the radius of the source and $\nu$ is a typical frequency
of the radiated spectrum. Again fixing $R$ to be the Schwarzchild
radius of the BH, we have
\begin{equation} n_{ph}=2\times 10^{15} L_{45} M_8^{-2}
\left( {1\ \hbox{keV}\over E_{ph}}\right) \
\hbox{photons/cm}^3,\end{equation}
so that the vorton mean free path
$\lambda_{_V}=(n_{ph}\sigma_{_V}^{_{Th}})^{-1}$ is
\begin{equation} \lambda_{_V}=2\times 10^{29} Z_{100}^{-2}
\left({m\over 10^9 m_p}\right)^2 M_8^{2} 
\left({E_{ph}\over 1\ \hbox{keV}}\right) L_{45}^{-1}\ \hbox{cm}
\end{equation}
which is much greater than the typical acceleration length $\ell_s
\simeq 3\times 10^{13} M_8$~cm.

In brief, the acceleration of vortons in the hot spots of
radio galaxies is more favorable than the same mechanism applied to
protons because synchrotron and Compton losses scale respectively like
$$Z^2 \left({m_p\over M_{_V}}\right)^4\simeq
10^{-22} Z_{100}^{-2}\ \ \hbox{(synchrotron)}$$
and
$$Z^4 \left({m_p\over M_{_V}}\right)^4
\simeq 10^{-18} Z_{100}^0 \ \ \hbox{(Compton)}.$$
In what follows, we postulate that vortons that are uniformly
distributed in the accretion disk around a BH can reach the region of
high electrostatic field. Taking into account Eq.~(\ref{ii}), we see
that they can gain energies of
\begin{equation} E_{_V}=10^{20} Z \hat a L_{45} \ \hbox{eV},\end{equation}
so that even active galaxies with luminosities as low as
$10^{43}$~erg/s can accelerate vorton to $10^{20}$~eV if the angular
momentum of the BH is large enough that $\hat a\sim 1$.

As already stated, acceleration at these energies of vortons can happen
also in the radio galaxy hot spots. The possibility of accelerating
them near the central BH means that under the hypothesis that
monopolar electrostatic and rotating BH is the power engine of all
active galaxies~\cite{begel}, Seyfert~I, II, QSO and AGN are potential
vorton acceleration sites, thereby tremendously increasing the number
of sites as compared to the proton case since there is no GZK
cutoff in their case, as we shall now see.

\section{Energy Considerations and Propagation.}\label{sec:III}

The flux of cosmic rays with energy higher than $10^{20}$ eV is
estimated of the order of
$10^{-20}$~cm$^{-2}\cdot$s$^{-1}\cdot$ster$^{-1}$, i.e. a flux
$10^{-19}$~cm$^{-2}\cdot$s$^{-1}$, so the energy flux of particles
having energies $E=E_{20}\times 10^{20}$eV is
\begin{equation} \Phi = 10^{-22} E_{20}\ 
\hbox{erg}\cdot\hbox{cm}^{-2}\cdot\hbox{s}^{-1}.
\end{equation}
If ${\cal P}_{_V}$ is the probability that a vorton interacts with the
atmosphere and if all the particles with energies greater than
$10^{20}$ eV are vortons, then the actual flux (as opposed to the
observed one) is
\begin{equation} \Phi = 10^{-6} E_{20}{\cal P}_{5}\
\hbox{erg}\cdot\hbox{cm}^{-2}\cdot\hbox{s}^{-1},
\end{equation}
with ${\cal P}_{5}={\cal P}_{_V} /10^{-5}$ (see Sec.~\ref{eff}). In order to
achieve this flux, extragalactic sources up to a distance $D$ must
supply a power $W$
\begin{equation} W=10^{47} E_{20} {\cal P}_{5} \left( {100\ \hbox{Mpc}\over
D}\right)^2 \ \hbox{erg}\cdot\hbox{s}^{-1}.\end{equation} Assuming
$10^5$ potential sources (Sy~I, I, QSO)~\cite{sources} means that each
source must supply a power
\begin{equation} W=10^{44} E_{20} {\cal P}_{5} \left( {1\ \hbox{Gpc}\over
D}\right)^2 \ \hbox{erg}\cdot\hbox{s}^{-1},\end{equation} power too
uncomfortably close to the averaged power radiated by active galaxies.
Note the neutrino hypothesis suffers even more of the same drawback.

Until now, we have assumed the typical mean free path of a $10^{20}$
eV vorton to be much larger than the Hubble radius, with the meaning
that a vorton can reach us, even if emitted at large redshifts. We
shall now prove that point by estimating the energy losses the vorton
has to experience on his way.

We start with synchrotron radiation due to intergalactic magnetic
field ${\bf B}$ since this is supposed to be the most efficient
mechanism. The energy loss per unit time is
\begin{equation} {dE\over dt} = {2\over 3} {(Ze)^4\over E^2} c\gamma^4
B^2,\end{equation} in a mean magnetic field $B=\langle {\bf B}^2
\rangle ^{1/2}$ if the Larmor radius of the vorton is less than 
the correlation length of the magnetic field ${\bf B}$. The typical
slow down timescale for the vorton is thus given by
\begin{eqnarray} t_{_V} &=& {3\over 2} {M_{_V}^4 c^7\over E \langle
{\bf B^2} \rangle (Ze)^4} \nonumber\\ &=& 4.6 \times 10^{59} \left(
{B\over 10^{-8}\hbox{gauss}}\right)^{2}
E_{20}^{-1} \left({m\over 10^9 m_p}\right)^4
\hbox{ s} \end{eqnarray} which is clearly sufficiently large that
there is not even a cosmological cutoff for vortons.

A similar conclusion holds for the energy losses due to the microwave
background. Note however the essential difference in this case between
protons and vortons: if protons were pointlike particles with no
internal quark structure, they would be able to propagate almost
freely in the microwave background, even at these energies, and they
decay dominantly because the binding energy between quarks and gluons
is much less than the proton mass. Vortons, on the other hand, not
only have energy levels which are much higher than protons, but also
they propagate with a relatively low velocity ($\gamma\sim 10^6$).
Therefore, vortons, once accelerated can arrive on earth with
undegraded energy: supposing they were accelerated at the time of
formation of the galaxies ($z\sim 1$)~\cite{hammer}, they would by now
have lost a mere factor of two because of cosmological redshift. The
interesting point in that observation is that for $z\sim 1$, the ratio
between active and normal galaxies is about
$0.1$~\cite{hammer,tresse}.  Keeping that in mind, we can now
calculate the high energy vorton density under the hypothesis that
vortons were accelerated at the birth of galaxies.

If one assumes that all the unexplained cosmic rays with energy larger
that $10^{20}$eV are made of vortons, their density will be
\begin{equation} n_{_V}={4\pi\over c} {\Phi_{_V} \over {\cal P}_{_V}}
=4.2 \times 10^{-25} \Phi_{20} {\cal P}_{5}^{-1} \
\hbox{V}\cdot\hbox{cm}^{-1},\end{equation}
where $\Phi_{_V}$ (and
$\Phi_{20}=\Phi_{_V}/10^{-20}$~cm$^{-2}\cdot$s$^{-1}\cdot$ster$^{-1}$)
is the high energy cosmic ray flux.

The total number of vortons $N_{_V}$ in a sphere of radius
corresponding to $z=1$ is
\begin{equation} N_{_V}=3\times 10^{60} h^{-3} {\Phi_{20}
\over {\cal P}_{5}},\end{equation}
and their energy is $10^{20}(1+z) N_{_V}$~eV~$=6\times 10^{68}
h^{-3} \Phi_{20}/{\cal P}_{5}$~ergs, where $h$ is the Hubble
constant $H_0$ in units of 75 km$\cdot$s$^{-1}\cdot$Mpc$^{-1}$. Such
an amount of energy can be released by a number of active galaxies
$N_{_G}$
\begin{equation} N_{_G} = {2\times 10^9\over \varepsilon} h^{-3}
{\Phi_{20}\over {\cal P}_{5}} T_9^{-1} L_{43},\end{equation} with
$T_9$ the duty time of the galaxies in millions of years and
$\varepsilon$ the ratio between the power used to accelerate vortons
and the electromagnetic luminosity; note this latter parameter is not
constrained and can in fact exceed unity.

The baryonic mass in the sphere with $z=1$ is $M_{_B}=5.8\times 10^{54}
h^{-1}$~g, so that the total number of galaxies is
\begin{equation} N_{_G}\simeq 3\times 10^{10} h^{-1} M_{10}^{-1},
\end{equation} where $M_{10}$ is the galaxy mass in $10^{10}$ solar masses.
From Refs.~\cite{hammer,tresse}, we know the ratio between active and
normal galaxies to be of order 10\%, so the required efficiency is,
expressed as a function of solar masses accreted by the BH per year
$\dot M$,
\begin{equation} \varepsilon = 10^{-4} h^{-2} {\Phi_{20}\over
{\cal P}_{5}}T_9^{-1} M_{10} \left({\dot M\over \dot
M_\odot}\right)^{-1}.
\end{equation}

Finally, let us compute the required number of vortons present in the
universe in the form of dark matter. Let $\alpha$ be the ratio between
the total vorton and baryon masses in the universe. The ratio $\eta$
between vorton and baryon number densities is
\begin{equation} \eta = \alpha {m_{_P}\over M_{_V}},\end{equation}
and assuming the baryon density to be 10\% of the critical density
($n_b = 6.6\times^{-7} h^{2}$~baryons/cm$^{3}$) we finally obtain
\begin{equation} \alpha = 6.4\times 10^{-4} {\Phi_{20}\over
{\cal P}_{5}}Z_{100} \left( {m\over 10^9 m_{_P}}\right).\end{equation}
This result is understandable in two ways. Either all vortons present
in AGNs are indeed accelerated, in which case vortons are, according
to this calculation, expected to represent roughly less than a
thousandth of the matter in the universe, or, conversely, vortons do
fill the universe so that the actual value of $\alpha$ should exceed
unity. In the latter case, our calculation reveals that either the
interaction probability is lower than what we evaluated above, or only
a small fraction of vortons get accelerated. In any case, it should be
clear that our model provides a very high energy cosmic ray flux which
can be made to agree with observations.

\section*{Conclusions}

We have exhibited a model for explaining extremely high energy cosmic
rays that have recently been observed. More of these events are
expected to be observed in the near future by the Auger
Observatory~\cite{auger}. We propose that they are bound states of
very massive particles in vortons, i.e. loops of superconducting
cosmic strings stabilized by a current, that are are freed by
inelastic collision with atmospheric hadrons. The model, in its
simplest form, has only one free parameter, namely the energy scale at
which the current sets up in cosmic strings. This could easily be
extended to take into account a possible difference between this mass
scale and the string energy scale itself~\cite{ring2}. In this
particular framework however, the mass scale is constrained by
requiring that vortons do not overfill the universe, a bound that is
almost saturated by the demand that high energy cosmic rays are made
of vortons. This interesting coincidence could imply, if verified,
that a non negligible fraction of the dark matter in the universe
consists of vortons.

Let us here summarize our findings concerning the model. First, as
mentioned earlier, it essentially has only one free parameter, namely
the mass scale at which the superconducting current sets up in the
core of the cosmic strings. As it turns out, demanding these objects
to be candidates for the few $10^{20}$~eV events through interaction
with atmospheric protons completely fixes this parameter to $10^9$~GeV.
There doesn't seem to be any way out of this prediction, which renders
the model very falsifiable.

Another point worth mentioning is that the vorton-proton (or neutron)
cross section is quite weak, giving, at these energies, a total
interaction probability between $10^{-5}$ and $10^{-4}$ with the earth
atmosphere. Note again that this probability depends on nothing but
the fixed energy scale, and that, in order to fit the observed data,
it implies a high energy vorton flux whose numerical value is also
therefore determined.

No high energy cosmic ray model is complete without specifying
acceleration and propagation processes. In our case, acceleration is
performed very simply by kicking the vortons (at least those which
have been ionized somehow beforehand) with the high electrostatic
fields that are expected to be present in AGNs. The required quantity of
energy turns out to exist in these objects and because of the relatively
low amount of acceleration required ($\gamma\alt 10^6$), losses
are negligible and the vortons can escape the acceleration zone. This
is, to the best of our knowledge, the most efficient mean of extracting
$10^{20}$~eV in a single particle out of any astrophysical object.

Similarly to the fact that energy losses are negligible in the region
of acceleration, and for the same reason, the propagation in
intergalactic medium is done almost without any collision, thus making
this medium effectively transparent to very high energy vortons. As a
result, there is no reason for a cosmological (GZK) cutoff and vortons
can come from as far as a redshift of a few. This is satisfactory
because the ratio of active to normal galaxies increases as one goes
farther away, and the final vorton flux we end up with thanks to this
fact is actually quite large and indeed requires only a very small
fraction of the total mass density in the universe to consist in
vortons (of the order of $10^{-4}$) if all vortons are ionized (and
therefore accelerated), or, if one believes the standard vorton
predictions~\cite{ring2} that for a mass scale of $10^9$~GeV they
should be the dominant part of the dark matter, then one just needs
the same tiny fraction of vorton to be accelerated. The absence of any
cutoff also means we predict a spatial isotropy together with a
correlation with active galaxies.

Interacting with the atmosphere with a very low probability (and then
presenting a threshold followed by a line spectrum, presumably
unobservable unfortunately), vortons are here predicted to give rise
to mostly horizontal air showers, just like the neutrinos. Let us
stress however that until now, it has not been possible to find any
astrophysical mechanism that would give neutrinos such high
energies. Hence, although our vortons are indeed hypothetical in the
sense that we don't know what is the actual theory that describes
physics beyond the electroweak scale at which we expect to find them,
it should however be clear that they do not, apart in their existence
itself, imply any new mechanism of any kind.

Finally it is worth pointing out that the acceleration mechanism we
have proposed is just one amongst a few other acceleration mechanisms
already proposed in the same context such as the Fermi
mechanism. Vortons can be accelerated as well across the shocks of the
jets of the radio galaxies. We prefer however the direct acceleration
mechanism because of its high efficiency and therefore the energy
budget constraints can be more easily fulfilled. Note that this is not
the case for the neutrinos: in fact, as already stated, $10^{20}$eV
neutrinos have interaction cross sections (with hadrons) that are
roughly equivalent to vorton's. Consequently, it is highly
questionable whether the energetic constraints required to accelerate
the parent charged particle of the neutrinos in the lobes of radio
galaxies can be satisfied.

To end up this conclusion, let us compare our expectations with those
of the Auger Observatory~\cite{auger}, a project specifically designed
to accumulate more statistics on cosmic rays with energies in excess
of $10^{19}$~eV. The Auger Observatory will have an energy resolution
less than 20~\% which is presumably going to be insufficient to
actually allow definite conclusion about the line features. In spite
of this difficulty, it is however not impossible that part of the line
signal might be found using correlation function techniques after a
power-law substraction will have been made in the data: no significant
feature should be left by almost all the competing candidates. Another
characteristic of the Auger Observatory concerns its ability to
identify the primary of an air shower through accurate measurement of
the shower maximum ($\Delta X_{max} \sim 20$g$\cdot$cm$^{-2}$). In a
bound state model such as the one we propose here, the line part of
the spectrum should be initiated by radiative decay of the excited
vortons, hence the primary should be a photon. Then once the continuum
is reached, some other particle (the $\Sigma$ in our model) plays the
part of the primary, initiating a shower whose maximum is expected
somewhere else. It is not clear yet, because very model dependent,
whether the Auger Observatory accuracy in this measurement will be
sufficient, but it is also a firm prediction of any bound state model.
Finally, according to our calculation, the interaction probability
with the atmosphere is very low, with the result that many horizontal
showers are expected. Contrary to most other cosmic ray detectors, the
Auger Observatory will be very efficient to detect those, being also
understandable as a neutrino detector with an acceptance of 10
km$^3\cdot$sr water equivalent for horizontal air showers at
$10^{19}$eV~\cite{murat}. Finally, although the angular resolution of
the two detectors (giant array and air fluorescence) in the Auger
Observatory should be of the order of less than $2^o$, some events
will be observed in hybrid mode by both with a resolution of $0.3^o$,
a precision that is expected to be enough to conclude on the isotropy
of the cosmic ray sources.

All these facts lead to the conclusion that our model has a very high
potential for being either confirmed or ruled out shortly after the
Auger Observatory will be started.

\section*{Acknowledgments}
We wish to thank E.~Audit, B.~Carter, A.C.~Davis, A.~Gangui,
T.~Kibble, J.~P.~Lasota and H.~Sol for many illuminating
conversations, as well as many people from the Pierre Auger Project,
in particular J.~Cronin and M.~Boratav who accepted to spend some time
explaining the project.

\end{document}